\documentclass[12pt]{article}

\usepackage{epsfig,amsmath,amssymb,latexsym,axodraw}

\providecommand{\beqa}{\begin{eqnarray}}
\providecommand{\eeqa}{\end{eqnarray}}

\providecommand{\Abar}{\bar{A}}

\providecommand{\Mbar}{\bar{M}}
\providecommand{\Nbar}{\bar{N}}

\providecommand{\abar}{\bar{a}}
\providecommand{\bbar}{\bar{b}}

\providecommand{\we}{\wedge}
\providecommand{\tr}{\text{tr}}
\providecommand{\sign}{\text{sign}}

\def\cO{{\cal{O}} }
\def\cS{{\cal{S}} }

\def\cF{{\cal{F}} }

\numberwithin{equation}{section}

\begin{document}

\thispagestyle{empty} \phantom{}\vspace{-1.5cm}
\rightline{HU-EP-03/42, UMD-PP-03-049}

\begin{center}
{\bf \large Enlarging the Parameter Space of Heterotic M-Theory}\\[3mm]
{\bf \large Flux Compactifications to Phenomenological Viability}
\end{center}

\vspace{0.6truecm}
\centerline{Gottfried Curio
$^{a,}$\footnote{curio@physik.hu-berlin.de} and Axel Krause
$^{b,}$\footnote{krause@physics.umd.edu}}
\vspace{.6truecm}

{\em \centerline{$^a$ Humboldt-Universit\"at zu Berlin,}
\centerline{Institut f\"ur Physik, D-12489 Berlin, Germany}}
\vspace{.3truecm}
{\em \centerline{$^b$ Department of Physics, University of
Maryland,} \centerline{College Park, MD 20742, USA}}

\vspace{0.6truecm}
%%%%%%%%%%%%%%%%%%%%%%%%%%%%%%%%%%%%%%%%%%%%%%%%%%%%%%%%%
%\vspace{.4truecm}
\begin{abstract}
\noindent Heterotic M-Theory is a promising candidate for that
corner of M-theory which makes contact with the real world.
However, while the theory requires one of its expansion
parameters, $\epsilon$, to be perturbatively small, a successful
phenomenology requires $\epsilon = {\cal O}(1)$. We show that the
constraint to have small $\epsilon$ is actually unnecessary:
instead of the original flux compactification background valid to
linear order in $\epsilon$ one has to use its appropriate
non-linear extension, the exact background solution. The exact
background is determined by supersymmetry and consequently one
expects the tree-level cosmological constant to vanish which we
demonstrate in detail, thereby verifying once more the consistency
of this background. Furthermore we show that the exact background
represents precisely the 11d origin of the 5d domain wall solution
which is an exact solution of the effective 5d heterotic M-theory.
We also comment on singularities and the issue of chirality
changing transitions in the exact background. The exact background
is then applied to determine Newton's Constant for vacua with an
M5 brane on the basis of a recent stabilization mechanism for the
orbifold length. For vacua without M5 brane we obtain a correction
to the lower bound on Newton's Constant which brings it in perfect
agreement with the measured value.
\end{abstract}

hep-th/0308202

\newpage
\pagenumbering{arabic}

\section{Introduction}

Heterotic M-theory \cite{HW1,HW2} is a prime candidate for
addressing successfully low-energy phenomenology like the
unification of all coupling constants\footnote{Alternatively Grand
Unification can be embedded into String-/M-Theory through
D-branes; see e.g.~\cite{GUT}.} \cite{WWarp}, predicting the
observed value for Newton's Constant \cite{WWarp}, getting soft
supersymmetry breaking terms of the right size \cite{SBT} (in
particular the gaugino masses) and  obtaining lepton and quark
mass hierarchies or neutrino masses of the right size \cite{ADH}.
Moreover, it presents a natural arena for any implementation of
hidden sector physics like supersymmetry breaking through hidden
sector gaugino condensation \cite{FGN,DIN,DRSW}.

The novel feature of heterotic M-theory was the necessity to deal
with backgrounds with non-vanishing vacuum expectation values for
its four-form field strength $G$. The reason being that in the
Bianchi-identity for $G$ magnetic sources given by the boundary
Riemann curvature two-forms together with the boundary super
Yang-Mills (SYM) curvature two-forms appear. Therefore, the
Bianchi identity cannot be solved anymore by embedding the spin in
the gauge connection (as in the weakly coupled heterotic string)
and requires a non-vanishing $G$ of order $\kappa^{2/3}$, where
$\kappa$ is the 11d gravitational coupling constant. However, a
nontrivial $G$ is related through the gravitino Killing spinor
equation typically to a warp-factor deformed background if one
wants supersymmetry to be preserved.

Therefore the challenge was to find backgrounds corresponding to
compactifications in the presence of $G$ fluxes which would break
the 16 supersymmetries of the theory exactly down to four,
corresponding to a 4d low-energy theory with desired N=1
supersymmetry. This was achieved in \cite{WWarp} for the case of a
deformed Calabi-Yau (CY) threefold compactification with
deformation controlled by the $G$ fluxes. For the case which is
interesting for phenomenology $G$ possesses a non-vanishing
component only along the CY, thereby inducing a variation of the
CY size along the orbifold direction (via the warp-factor/$G$-flux
balance from the Killing spinor equation). Notice that this
phenomenon has no counterpart in the weakly coupled heterotic
string because the required flux $G_{lmnp}$ ($l,m,\hdots$ denote
the CY indices) is projected out in that limit in consistency with
the fact that the heterotic string's NS field-strength $H_{mnp}$
originates in M-theory from $G_{mnp 11}$. Therefore the varying CY
volume is a particular feature of the strongly coupled heterotic
string.

For the case with non-vanishing $G_{mnpq}$ but vanishing $G_{mnp
11}$ the background geometry describing the compactification of
11d heterotic M-theory down to 4d found in \cite{WWarp} turned out
to be the following warped geometry ($\mu,\nu,\hdots$ refer to the
external Minkowski spacetime indices, $g_{mn}$ is the CY metric
and $x^{11}$ denotes the $\mathbf{S}^1/\mathbf{Z}_2$ orbifold
coordinate)
\beqa
ds^2 = (1-f_{lin}(x^{11}))\eta_{\mu\nu}dx^\mu dx^\nu
+ (1+f_{lin}(x^{11}))\big(g_{mn}(y)dy^m dy^n +(dx^{11})^2\big)
\label{WittenSol}
\eeqa
with warp-factor $f_{lin}$ given by ($\omega$ denotes the CY
K\"ahler-form)
\beqa
f_{lin}(x^{11}) = x^{11} \frac{2\pi}{3 V_v}
\big(\frac{\kappa}{4\pi}\big)^{\frac{2}{3}}
\int_{CY}\omega \we \frac{(\tr F\we F - \frac{1}{2}\tr R\we
R)_v}{8\pi^2} \; .
\label{LinWarp}
\eeqa
The integral represents the $G$ flux on the visible boundary and
$V_v$ is the visible boundary's CY volume. If the visible
`instanton number' is larger than the hidden one the integral
gives a negative value \cite{WWarp}. Without loss of generality we
will therefore designate the visible boundary the one with larger
`instanton number' so that $f_{lin}$ can always be regarded as
being non-positive.

An important point is the regime of validity of the background
(\ref{WittenSol}). Its derivation \cite{WWarp} assumed an
expansion in the small ``parameter'' $f_{lin}$ -- the warp-factor
-- and kept all terms linear in $f_{lin}$. It therefore looses
validity when the neglected higher order contributions
$f_{lin}^n,\; n\ge 2$ become of the same size as $f_{lin}$.
Because of the linearity in $x^{11}$ the expansion `parameter'
$f_{lin}$ will however grow until at some finite critical distance
$x^{11}=L_c$ the factor $1+f_{lin}$ which multiplies the internal
geometry becomes zero
\beqa
1+f_{lin}(L_c) = 0  \; .
\label{critical bound}
\eeqa
One consequently has to impose $L_c$ as an upper bound on the
orbifold length $L$ (distance between visible and hidden boundary)
in order to avoid an unphysical regime of negative metric with
negative internal volume beyond this critical distance. Moreover,
in the vicinity of the critical distance $f_{lin}(x^{11}\lesssim
L_c)$ necessarily becomes of order one instead of staying
perturbatively small as would be required for the validity of
(\ref{WittenSol}). Therefore the linearized background
(\ref{WittenSol}) actually breaks down before one has reached the
critical distance and the inclusion of the neglected higher order
contributions $f_{lin}^n, \; n\ge 2$ becomes mandatory.

A criterion for the validity of (\ref{WittenSol}) which is
ubiquitous in the literature comes as follows. By plausibly
estimating the absolute value of the integral in (\ref{LinWarp})
to be of order $V_v^{1/3}$ (where one usually equates in addition
the compactification scale $V_v^{1/6}$ with the inverse of the
Grand Unification (GUT) scale $1/M_{GUT}$) one obtains that
$|f_{lin}|$ is bounded from above by
\beqa
\epsilon
= \frac{2\pi}{3V_v^{2/3}}
\Big(\frac{\kappa}{4\pi}\Big)^{\frac{2}{3}}L
\ge \frac{2\pi}{3V_v^{2/3}}
\Big(\frac{\kappa}{4\pi}\Big)^{\frac{2}{3}}x^{11}
\simeq |f_{lin}(x^{11})|
\; ,
\label{Epsilon}
\eeqa
where $L$ has to be smaller than $L_c$ as explained before.
Because (\ref{critical bound}) means $|f_{lin}(L_c)|=1$, this
upper bound on $L$ translates to the upper bound $\epsilon \leq 1$
(as already an estimate went into (\ref{Epsilon}) the bound is
rather $\epsilon \leq {\cal O}(1)$). However, this constraint is
still not restrictive enough to guarantee the validity of
(\ref{WittenSol}). Namely, to make sure that the higher order
corrections $f_{lin}^n, \; n\ge 2$ are sufficiently suppressed,
one has to sharpen the bound to \cite{BD1}
\beqa
\epsilon \ll 1   \; .
\label{EpsCon}
\eeqa
Naturally, the effective 4d action \cite{LOW1} which results from
the dimensional reduction of 11d heterotic M-theory over the
linearized background (\ref{WittenSol}) thus inherited the
requirement for small $\epsilon$.\footnote{Actually another
dimensionless expansion parameter $\epsilon_L = R_v/L$ (where $R_v
= V_v^{1/6}$) appears in the effective 4d theory. However, the
background (\ref{WittenSol}) is exact in this parameter to all
orders \cite{OvrutLec} which is why this parameter is of no
particular concern to us here. By requiring $R_v,L \gg
\kappa^{2/9}$ in addition one is assured that supergravity is a
good approximation to M-theory and that all corrections due to
geometrical instantons are sufficiently suppressed.} It had been
realized early on, however, that the regime where $\epsilon$ is
small is hardly compatible with the requirements of phenomenology
which demands $\epsilon\simeq\cO(1)$ \cite{BD1}. Moreover, one has
to be careful with the 4d Newton's Constant which becomes
negative, as a result of the negative internal volume, when
$\epsilon$ becomes too large. More precisely, one finds the tree
level relations \cite{BD2} (here $R_v = V_v^{1/6}$)
\beqa
L^2 = \frac{\alpha_{GUT}^3V_v}{512\pi^4 G_4^2} \; , \qquad
\kappa^{2/9} = R_v(2(4\pi)^{-2/3}\alpha_{GUT})^{1/6}
\eeqa
which with $M_{GUT}=1/R_v=3\times 10^{16}$GeV, $\alpha_{GUT}=1/25$
and the experimental value for Newton's Constant give
\beqa
L = 12 \kappa^{2/9}\; , \qquad R_v = 2 \kappa^{2/9}
\eeqa
and therefore lead to an $\epsilon = \frac{(4\pi)^{1/3}}{6}
\frac{L/ \kappa^{2/9}}{(R_v/ \kappa^{2/9})^4}={\cal O}(1)$. Notice
the sensitivity due to the fourth power to the actual value for
$M_{GUT}$ which might range between $2\times 10^{16}$GeV and
$3\times 10^{16}$GeV. Thus -- apart from possible corrections to
the estimate $V_v^{1/3}$ for the flux integral -- this generic
discrepancy, that the ``real world'' seems to reside where the
hitherto available effective theory essentially breaks down
presents a major obstacle for any full-fledged contact between
M-theory in its heterotic corner and phenomenology.

It is the aim of this paper to show that actually the small
$\epsilon$ constraint (\ref{EpsCon}) of heterotic M-theory becomes
dispensable and therefore the discrepancy between theory and
phenomenology disappears when one uses the proper non-linear
extension \cite{CK1} to the linear background (\ref{WittenSol})
which includes the higher-order corrections $f^n_{lin}$ neglected
in (\ref{WittenSol}). By using this non-linear background for a
reduction of the 11d theory also the 4d effective heterotic
M-theory would become exempt from the small $\epsilon$ constraint
which would open up the way to reliably obtain phenomenological
predictions from M-theory in its heterotic regime.

To lend further support to our inclusion of non-linear terms in
the derivation of the exact background we connect this background
to two exact solutions already considered in the literature.
First, concerning the case of compactification of heterotic
M-theory to six dimensions, an exact solution was presented
already by Witten in \cite{WWarp}. There, precisely as in our
case, one had a flux source of first-order in $\kappa^{2/3}$
together with an exact (i.e.~non-linearly extended) warp-factor,
the latter as in our case demanded by supersymmetry (these issues
are explained in detail in subsection 2.4). Secondly, our exact
background is actually the 11d origin of the 5d domain wall
solution discovered in \cite{DW} (a supersymmetric solution to the
equations of motion of the effective 5d heterotic M-theory), which
being an exact solution also incorporates higher order
$\kappa^{2n/3},\;n>1$ contributions; furthermore the linear
truncations correspond also to each other under dimensional
reduction.

The remaining organization of the paper is as follows. In section
two we first review the non-linear compactification background and
point out why it exempts the theory from its small $\epsilon$
constraint. Next, we discuss for the case where an isolated
singularity occurs the issue of chirality change for the hidden
boundary gaugino once the singularity is passed. Here we find
compelling reason why the hidden boundary should be placed at or
before (in the orbifold direction) the singularity. We then show
in detail the relation between the linear and the non-linear
background, in particular how the latter incorporates all the
higher order corrections which lead to the absence of the small
$\epsilon$ constraint (and point to the 6d analogy). As an example
we describe the non-linear background for the simplest heterotic
M-theory vacua without additional M5 branes where a singularity
appears at some finite $x^{11}_0$ along the orbifold whose value
is determined by the visible boundary flux. In section three we
demonstrate the connection with the 5d domain wall solution. In
section four we show that the heterotic M-theory tree-level
cosmological constant vanishes for the exact non-linear
background. This provides an additional verification that the
structure of the non-linear background which includes field theory
corrections of all $\kappa^{2n/3},\; n\ge 1$ orders is indeed
protected by supersymmetry. In section five we use the non-linear
background to derive the 4d Newton's Constant $G_4$. For the
simplest vacua without M5 branes in case that there is no
M-theoretic singularity resolution or that the resolution affects
the geometry only locally, one has to place the hidden boundary
before the singularity which leads to a lower bound for $G_4$.
With the numerical input of the GUT scale and the GUT gauge
coupling this lower bound is in excellent agreement with the
measured value. Furthermore, we evaluate Newton's Constant also
for vacua with an additional parallel M5 brane which shows that
for the stabilization scenario considered in \cite{CK2} $G_4$ is
very close to the measured value.

\section{The Exact Non-Linear Flux Compactification Background of
Heterotic M-Theory}

We saw before that for the linearized background (\ref{WittenSol})
one has to cut-off space at a critical orbifold distance $L_c$ but
to ensure that $\epsilon\ll 1$ requires actually a much smaller
cut-off in $L$. It seems that there are two ways how one might
deal with such a situation. Either one is able to find a
stabilization mechanism for the orbifold size modulus (in a
strongly coupled disguise this is nothing else but the dilaton
stabilization problem) which allows to stabilize the orbifold size
at a sufficiently small value of $L$ (below $L_c$) or one has to
find the extension of (\ref{WittenSol}) which presumably avoids
negative warp-factors and CY volumes at all.

The first possibility was addressed in \cite{CK2} where a
stabilization mechanism\footnote{An alternative moduli
stabilization mechanism for M-theory was recently proposed in
\cite{BSA} in the context of M-theory compactifications on $G_2$
holonomy manifolds. This mechanism gives a stabilized vacuum which
is supersymmetric and possesses a negative cosmological constant.}
through non-perturbative open membrane instanton effects in
conjunction with $G$-fluxes was found in the presence of an
additional parallel M5 brane which fills the external spacetime
and wraps an internal holomorphic 2-cycle (see also \cite{MPS} for
earlier important insights related to this set-up and
\cite{LOPR}). Though a local minimum of the moduli
potential\footnote{Earlier runaway boundary-boundary potentials
were derived e.g.~in \cite{KPot}.} could be established at
positive vacuum energy, there are general arguments \cite{deSitt}
that such de Sitter vacua should be false vacua and the global
potential should at least allow for another zero energy-density
vacuum in the decompactification limit. Therefore, even if one is
able to stabilize the orbifold modulus below $L_c$ there are
nevertheless compelling reasons why one would like to study the
moduli potential globally, i.e.~at best for arbitrary values of
$L$. So either way one has to understand how (\ref{WittenSol}) has
to be extended which we will now address.

\subsection{The Exact Non-Linear Flux Compactification Background}

In \cite{CK1} (cf.~also \cite{K1}) the relation between an 11d
heterotic M-theory compactification background preserving 4d N=1
supersymmetry
\beqa
ds^2 = e^{b(y,x^{11})}\eta_{\mu\nu} dx^\mu dx^\nu
+ e^{f(y,x^{11})}g_{mn}(y) dy^m dy^n
+ e^{k(y,x^{11})}dx^{11} dx^{11}
\label{WarpMetric}
\eeqa
and the four-form flux $G$ was investigated. As the Ansatz for the
metric reflects, it was assumed that heterotic M-theory is
compactified on a seven-space which is a conformal warp-factor
deformation of the $\text{CY}\times\mathbf{S}^1/\mathbf{Z}_2$
geometry.

The non-trivial four-form flux $G$ which arises in heterotic
M-theory e.g.~by the presence of its boundaries implies a
non-trivial Bianchi identity
\beqa
dG = \sum_i \delta(x^{11}-x^{11}_i)S_i(y)\we dx^{11}
\label{BI}  \; ,
\eeqa
where the $S_i(y)$ four-form describes the magnetic sources --
boundaries and possible additional parallel M5 branes. It turns
out that one can solve the gravitino Killing-spinor equation
explicitly and thus determine the supersymmetry preserving
compactification backgrounds for arbitrary sources $S_i$ as long
as they are localized along the orbifold and thus do not depend on
$x^{11}$ \cite{CK1}. In particular the $S_i$ could therefore
accomodate higher order corrections as long as these corrections
do not change the formal structure of the 11d gravitino
supersymmetry transformation. However, the higher order
$\kappa^{4/3}$ corrections which are needed in heterotic M-theory
in order to smooth out the localizing delta-functions and thus to
provide the boundaries with a finite thickness are beyond the
grasp of the 11d supergravity framework. Here, the 10d boundary
$E_8$ super Yang-Mills fields would have to propagate into the 11d
bulk. However, supersymmetric Yang-Mills theories in the
conventional local field-theory framework are impossible in
spacetime dimensions higher than ten \cite{Nahm}. Therefore as
long as the description in terms of supergravity is adequate we do
not expect a smoothening of the localizing delta function in the
Bianchi identity (\ref{BI}). Subsequently we will work in
Horava-Witten supergravity at $\kappa^{2/3}$ order where these
higher order `quantum M-theory' corrections are absent.

The magnetic sources $S_{v,h}(y)$ coming from the visible
resp.~hidden boundary are in Horava-Witten supergravity given
through \cite{HW2}
\beqa
S_{v,h}(y) = -\frac{1}{2\sqrt{2}\pi} ( \tr F\we F - \frac{1}{2}\tr
R\we R )_{v,h}
\left(\frac{\kappa}{4\pi}\right)^{2/3}  \; .
\label{HWBoundFlux}
\eeqa
Moreover, compatible with the 11d N=1 supersymmetry of heterotic
M-theory, there can also be M5 branes parallel to the boundaries
acting as magnetic sources for $G$. For instance an M5 brane which
is space-time filling in the 4d external directions and wraps an
internal holomorphic 2-cycle $\Sigma$ (with Poincar\'e dual
four-form $[\Sigma]$) gives a contribution
\beqa
S_{5}(y) = -\frac{4\pi}{\sqrt{2}}\left[\Sigma\right]
\left(\frac{\kappa}{4\pi}\right)^{2/3} \; .
\label{HWM5Flux}
\eeqa
For these two different types of magnetic sources one can solve
the Bianchi identity for the field-strength $G$ and obtains two
types of contributions, $G_{lmn11}$ and $G_{lmnp}$. In addition,
for the Bianchi identity to have a solution the anomaly
cancellation condition
\beqa
\sum_i S_i(y) = S_v(y)+S_h(y)+\sum_{M5's}S_5(y) = 0
\eeqa
has to be satisfied. For these two types of $G$ together with the
Ansatz (\ref{WarpMetric}) one can then solve the gravitino Killing
spinor equation, thus searching for 4d N=1 supersymmetry
preserving compactification geometries.

Let us state the result \cite{CK1}. Without loss of generality one
may equate the two internal warp-factors
\beqa
k(y,x^{11}) = f(y,x^{11}) \; .
\eeqa
For the two remaining warp-factors $b$ and $f$ one finds equations
which specify the warp-factors in terms of $G$ and are given as
the following first order partial differential equations ($a,\abar
= 1,\hdots,3$ refers to the holomorphic coordinates $y^a,{\bar
y}^{\bar a}$ on the CY)
\begin{alignat}{3}
-2\partial_a b &= \partial_a f =
i\frac{\sqrt{2}}{3}e^{-\frac{3}{2}f}\omega^{lm}G_{alm11}
\label{FirstSet} \\
-\partial_{11} b &= \partial_{11} f =
-\frac{\sqrt{2}}{24}e^{-\frac{3}{2}f}
\omega^{lm}\omega^{np}G_{lmnp}
\label{SecondSet}
\end{alignat}
(in the exponential on the right hand side (rhs) $k$ has already
been replaced by $f$). Therefore
\beqa
\partial_a\partial_{11}b = \partial_a\partial_{11}f = 0
\eeqa
which is solved in general by
\beqa
b = b_1(y) + b_2(x^{11}) \; , \qquad f = f_1(y) + f_2(x^{11}) \; .
\eeqa
One is therefore naturally led to consider two choices in detail.
One choice, where $G$ possesses only a component with index in the
orbifold direction
\beqa
G_{lmn11} \ne 0\; , \quad G_{lmnp}=0
\eeqa
leads to
\beqa
b=b(y)\; , \quad f=f(y) \; , \quad b(y)=-\frac{1}{2}f(y)
\eeqa
and can be used to show \cite{CK1} how the M-theory relation
between warp-factors and flux (\ref{FirstSet}) reproduces the
analogous one of the weakly coupled heterotic string with torsion
\cite{Strom1}. The other choice where $G$ is non-vanishing only on
the CY
\beqa
G_{lmn11}=0\; , \quad G_{lmnp} \ne 0
\label{SecFlux}
\eeqa
gives
\beqa
b=b(x^{11})\; , \quad f=f(x^{11}) \; , \quad b(x^{11})=-f(x^{11})
\label{SecChoice}
\eeqa
and leads to the phenomenologically interesting case with varying
CY volume along the orbifold direction. It is this latter case
which will be of main interest to us here and on which we will
focus exclusively during the rest of this paper.

For this second choice the only remaining nontrivial differential
equation (\ref{SecondSet}) for $f$ can easily be integrated to
\beqa
\label{integrated}
e^{\frac{3}{2}f(x^{11})} = 1 -
\frac{1}{8\sqrt{2}}\omega^{lm}\omega^{np}\int_0^{x^{11}}dx'^{11}
G_{lmnp}(y,x'^{11})
\eeqa
(we have set $f(0)=0$ meaning that the six-space compactification
geometry reduces to the undeformed CY when one approaches the
visible boundary). Notice that the rhs can become negative. In
this case the left hand side (lhs) will be defined by the rhs
through analytic continuation. To comprise also this case, let us
denote henceforth $e^{f(x^{11})/2}$ together with its analytically
continued negative values by $\cF(x^{11})$ whose cube is defined
through the rhs of (\ref{integrated}). Because the warp-factor
$e^f=\cF^2$ which enters the metric is always non-negative, this
analytic continuation indeed makes sense. We will comment on its
physical implications in the next subsection. It is the feature
that the sources of the theory are localized along the orbifold
which will now allow us to perform the integral explicitly.
Namely, with sources localized in the $x^{11}$ direction and
absence of any components $G_{lmn11}$ the Bianchi identity
(\ref{BI}) gets solved by
\beqa
G = \sum_i\Theta(x^{11}-x^{11}_i)S_i(y)
\eeqa
such that the $x^{11}$ and the $y$ dependence become decoupled and
the integration can be carried out explicitly with the result
\beqa
\cF^3(x^{11}) = 1 - \sum_i
(x^{11}-x_i^{11})\Theta(x^{11}-x_i^{11})\cS_i  \; .
\label{WarpFactorSol}
\eeqa
Here it is convenient to define the scalars
\beqa
\cS_i = \frac{1}{8\sqrt{2}}\omega^{lm}(y)\omega^{np}(y)
\left(S_i(y)\right)_{lmnp}
\eeqa
where the index contractions are done with the undeformed CY
metric such that there are no warp-factors present in the
contractions.

Using the partial differential equations for $f$ the $\cS_i$ are
seen to be independent of the CY coordinates $y$ and are therefore
constant. This allows to express them alternatively as flux
integrals
\beqa
\cS_i = \frac{1}{V_v}\int d^6y\sqrt{\det(g_{lm})}\cS_i
= \frac{1}{2\sqrt{2}V_v}\int_{CY} \omega\we S_i \; .
\eeqa
The anomaly cancellation condition can be restated as
\beqa
\sum_i \cS_i(y) = \cS_v(y)+\cS_h(y)+\sum_{M5's}\cS_5(y) = 0
\label{ACC}
\eeqa

\subsection{Non-Negative Metric and CY Volume}

Let us now explore the consequences of the solution
(\ref{SecChoice}), (\ref{WarpFactorSol}). First of all, we gain
from (\ref{WarpFactorSol}) the warp-factor
\beqa
e^{f(x^{11})} = \cF^2(x^{11}) = \Big|1 -
\sum_i(x^{11}-x_i^{11})\Theta(x^{11}-x_i^{11})\cS_i
\Big|^{2/3}
\label{WarpFactor}
\eeqa
and together with (\ref{SecChoice}) the metric background solution
\begin{alignat}{3}
ds^2 &= {\hat g}_{MN} dx^M dx^N \notag \\
&= e^{-f(x^{11})}\eta_{\mu\nu} dx^\mu dx^\nu
+ e^{f(x^{11})}g_{mn}(y) dy^m dy^n
+ e^{f(x^{11})}dx^{11} dx^{11} \; .
\label{MetricSol}
\end{alignat}
Evidently, the warp-factor and metric are now manifestly
non-negative in contrast to the linearized case
(\ref{WittenSol}),(\ref{LinWarp}).

There are two situations which can now arise: either the fluxes
and the distribution of the sources are such that the warp-factor
stays positive for arbitrary values of $x^{11}$ or it becomes zero
at some point. An example for the first case would be a vacuum
with an M5 brane whose flux compensates the visible boundary flux
and in the extreme case where these fluxes are equal but opposite
can lead to a constant and positive warp-factor at $x^{11}\ge
x^{11}_5$ where $x_5^{11}$ is the position of the M5 along the
orbifold (see e.g.~\cite{CK2}, fig.2 which uses the geometry
(\ref{MetricSol}) for this case). In the second case when the
warp-factor becomes zero at some point we run into a naked
singularity. That this is not merely a coordinate singularity can
be seen e.g.~by evaluating the Riemann scalar for the metric
(\ref{MetricSol}) (our general relativity conventions are those of
Weinberg \cite{Wein})
\beqa
R({\hat g}) = e^{-f}\big( 2f''+\frac{5}{2}(f')^2 \big)
\eeqa
which is easily seen to diverge at the point where $e^f$ becomes
zero. In a pure classical gravity framework this singularity would
have to be cut out of spacetime. However, because we are working
here in M-theory one should expect that the singularity gets
resolved by going beyond the tree-level supergravity
approximation. For instance higher order in derivative corrections
or M5 instanton contributions which wrap the total vanishing CY
volume might be able to resolve the time-like singularity. We will
however present evidence below which suggests that the hidden
boundary should sit at the singularity thus pointing to a possible
role gauge instantons could play in a resolution of the
singularity \cite{WWarp}. We will leave the detailed investigation
of the resolution mechanism to future work but want to stress that
such a resolution is hardly conceivable in the linearized
background where the CY volume and the warp-factor become
increasingly negative beyond the singularity.

Let us now see what happens to the restrictions on $\epsilon$
which were found in the linearized case. First of all as we will
show in detail in the next subsection, the exact background
corresponds to the non-linear extension of the linearized
background (\ref{WittenSol}) in the sense that it comprises all
higher-order corrections $f_{lin}^n, \; n\ge 2$ necessary for it
to be an exact solution. Consequently, this {\it eliminates the
sharper bound} $\epsilon\ll 1$ for the exact background whose very
task it was to make sure that these higher-order corrections do
not have to be included. Concerning the second milder restriction
of $\epsilon\le 1$ which came in the linearized background from
the cut off in $L$ in order to avoid a negative metric, the
crucial question will be whether one can resolve the singularity
by M-theoretic corrections or not. Under the premise that it can
be done this bound would also disappear because the remaining
geometry except for the singularity is well-behaved. However, we
will see in the next subsection that if the hidden boundary is
placed in the region beyond the singularity a chirality change for
the hidden boundary gaugino would occur and moreover the kinetic
term of this gaugino would appear with the wrong sign. The latter
seems to indicate that even if there is a local (in the $x^{11}$
sense) singularity resolution mechanism which leaves the global
structure of the geometry intact, nevertheless the hidden boundary
should sit at the singularity (it could also sit before the
singularity but the singularity position is distinguished in vacua
without M5 branes by leading to an astonishingly precise value for
Newton's Constant together with the usual GUT framework, cf.~the
last section). Consequently for the simplest vacua without
additional M5 branes the ensuing upper bound on $L$ is such that
still one would obtain a constraint $\epsilon \lesssim 1$ however
this time for very different reasons. Nevertheless, in either case
we see that when one wants to address heterotic M-theory
phenomenology which requires $\epsilon\simeq \cO(1)$ (see the
introduction) one should use the exact non-linear background
rather than the linearized one in order to avoid inconsistencies
with the sharper bound $\epsilon\ll 1$.

With explicit knowledge of the warp-factor, we can also determine
the CY volume as
\begin{alignat}{3}
V(x^{11}) &= \int d^6y\sqrt{\det(e^f g_{lm})} = |\cF|^6\int
d^6y\sqrt{\det(g_{lm})} = e^{3f}V_v
\notag \\
&= \Big( 1 - \sum_i \cS_i\times
(x^{11}-x_i^{11})\Theta(x^{11}-x_i^{11}) \Big)^2 V_v
\label{CYVol}
\end{alignat}
thus revealing as well a manifestly positive and in addition
simple quadratic behaviour. We will make the connection to the
linearized background (\ref{WittenSol}),(\ref{LinWarp}) precise
later.

\subsection{Chirality Change for the Hidden Gaugino}

Let us now analyze in more detail for the case in which a
singularity occurs what happens behind (in the $x^{11}$ direction
sense) the singularity. For this we will assume that M-theory
provides us with a resolution of the singularity such that the
resolution will only affect the local vicinity of the singularity
but leave the global geometry intact. Notice that without this
assumption the excision of the singularity would leave us with two
disconnected pieces of spacetime and thus effectively ending
spacetime (as it can be reached from the visible boundary) at the
singularity.

While the warp-factor $e^f$ is always non-negative, its
square-root $\cF(x^{11})$ becomes negative in the regime where
either the fluxes or $x^{11}$ grow large. For the simplest
situation with just the boundary sources being present the
singularity would sit at $x_0^{11}=1/\cS_v$. First a technical
remark: because beyond this point the right hand side of
(\ref{WarpFactorSol}) becomes negative, we cannot simply evaluate
$\sqrt{e^f}$ as $e^{f/2}$ (which is only valid for $x^{11}\le
x_0^{11}$) but must use its analytically continued value which
gives
\beqa
\sqrt{e^f} = |\cF| \; .
\eeqa
This generalization comprises the standard result
$\sqrt{e^f}=e^{f/2}$ whenever $\cF$ is positive. It means that a
quantity like $\sqrt{\det(e^f g)}$ on a space with odd dimensions,
e.g.~on the full internal 7-space, is given by $|\cF|^7\sqrt{\det
g}$ where the absolute value becomes essential in the region where
$x^{11}\ge x_0^{11}$.

Physically, the movement of the hidden boundary through the
singularity is related to a chirality changing transition (with
respect to movements in the length modulus $L$) for the gaugino on
the hidden boundary. The occurrence of such transitions in
string-theory was first pointed out in \cite{KS} (for some later
developments cf.~\cite{Cchiral}, \cite{MPS}). By following the
analysis in \cite{CK1} one sees that also the vielbein
${e^M}_{\Mbar}$ ($M,N,\hdots = 0,1,\hdots,9,11$ and $A,B,C,\hdots
= 0,1,\hdots,9$ in the following) is analytically continued when
we use the more general $\cF$ instead of $e^{f/2}$. For instance
the lhs of (\ref{integrated}) is derived from $e^{k/2-2f}$ which
stems from the vielbein combination $({e^{11}}_{\bar {11}})^{-1}
{e^l}_{\bar l} {e^m}_{\bar m} {e^n}_{\bar n} {e^p}_{\bar p}$
(contracted with $G_{lmnp}\eta^{{\bar l}{\bar m}} \eta^{{\bar
n}{\bar p}}$). Instead of the naive ${e^\mu}_{\bar\mu} = e^{f/2},
\; {e^m}_{\bar m} = {e^m}_{\bar m}(CY)\,e^{-f/2} ,\;
{e^{11}}_{\bar {11}} = e^{-f/2}$ (which are only valid for
$x^{11}\le x_0^{11}$) the vielbein becomes
\beqa
{e^\mu}_{\bar\mu} = \cF \; ,
\qquad {e^m}_{\bar m} = {e^m}_{\bar m}(CY)\,\cF^{-1} \; ,
\qquad {e^{11}}_{\bar {11}} = \cF^{-1}
\eeqa
which means it will likewise be continued to negative values in
the regime beyond the singularity. Again, the metric stays
positive under the continuation because it involves always two
vielbeine ${\hat g}_{MN} = {e_M}^{\Mbar} {e_N}^{\Nbar}
\eta_{\Mbar\Nbar}$. Now the chirality operator (from the 10d
perspective) is given by
\beqa
\Gamma^{11}(x^{11}) = {e^{11}}_{\bar {11}}(x^{11})\Gamma^{\bar
{11}}
= \cF^{-1}(x^{11})\Gamma^{\bar {11}}
\eeqa
with $\Gamma^{\bar {11}}$ the flat-space chirality operator. This,
however, means that the chirality operator which acts on the
hidden boundary $E_8$ gaugino $\chi$ as $\Gamma^{11}\chi =
\cF^{-1}\chi$ flips its sign once the hidden boundary passes the
singularity. In other words the chirality of the hidden gaugino
depends on whether the hidden boundary is placed before (positive
chirality) or behind (negative chirality) $x_0^{11}$.

Let us comment on coordinate transformations in this respect.
{}From the coordinates used in (\ref{MetricSol}) it seems that at
$x_0^{11}$ the orbifold itself would shrink to zero size. This,
however, is not true as a simple coordinate transformation from
$x^{11}$ to $y^{11}$ given by $dy^{11}=|\cF|dx^{11}$ shows.
Therefore, there is no shrinking of the orbifold size in $y^{11}$
coordinates -- instead the ${\hat g}_{11,11}$ part of the metric
stays constant. One could now think that such a coordinate
transformation might also eliminate the chiral transition because
it basically trivializes the vielbein in the orbifold direction.
Notice however, that the chiral transition is still present as we
remain after the coordinate transformation still with a discrete
sign factor which jumps at the singularity
\beqa
\Gamma^{11}(y^{11}) = |\cF| \Gamma^{11}(x^{11})
= \text{sign}(\cF)\Gamma^{\bar {11}} \; .
\eeqa
Because this chirality transition is caused by a sign change of
the vielbein it will however at the same time change the sign of
the 10d Dirac-operator $D\!\!\!\!\slash = {e^A}_{\Abar}
\Gamma^{\Abar} D_A$. Therefore, by placing the hidden boundary
beyond the singularity the gaugino would appear with a wrong sign
kinetic term. This strongly suggests that on physical grounds the
hidden boundary should be placed before or at the singularity.
Indeed, as we will see in the last section, the singularity
position stands out by leading to a very accurate agreement with
the measured Newton's Constant while incorporating all the
successes of the GUT theories.

\subsection{First Order Fluxes Do Not Imply Linearized Background}

Let us now turn to the relation between the non-linear background
and the linearized one. This of course poses the question `does
not working with order $\kappa^{2/3}$ sources already imply that
one has to use the linearized metric background?'. We will clarify
why this is not the case.

First the reader should notice that for the compactification of
heterotic M-theory on $K3\times \mathbf{S}^1/\mathbf{Z}_2$ down to
6d an exact solution was already found by Witten in \cite{WWarp}.
This background solution reads
\beqa
ds^2 = \big(c+w(x^{11})\big)^{-1/3} \eta_{\mu\nu} dx^\mu dx^\nu
+ \big(c+w(x^{11})\big)^{2/3} \big( g_{mn}(y) dy^m dy^n
+ dx^{11} dx^{11} \big)
\label{6dMetricSol}
\eeqa
where $c$ is a constant, $V_{K3}$ the K3 volume and in the limit
where the orbifold length $L$ is much bigger than the K3 radius
one obtains
\beqa
w(x^{11})\propto \frac{(L-x^{11})}{V_{K3}}\kappa^{2/3}
\int_{K3}\big(\tr F\we F-\frac{1}{2}\tr R\we R\big) \; .
\label{SourceParam}
\eeqa
As for the case of compactification to 4d we see also here that
$w(x^{11})$ contains a $G_{lmnp}$ flux of order $\kappa^{2/3}$. If
therefore working with $\kappa^{2/3}$ fluxes would necessarily
imply that the metric can never incorporate $\kappa^{2n/3},\,n>1$
contributions (after expanding the warp-factors) then one would
have to discard the exact solution (\ref{6dMetricSol}) and allow
only the truncated version where both warp-factors get linearized
to $c-\frac{1}{3}w(x^{11})$ resp.~$c+\frac{2}{3}w(x^{11})$. This
is however wrong as the full warp-factors of the exact solution
are demanded by 11d supersymmetry irrespective of what the actual
value of $G_{lmnp}$ is. The point here is that the exact solution
is a bulk solution expressed in terms of a largely unspecified
$G_{lmnp}$ flux. What matters to derive the exact solution is only
where $G_{lmnp}$ is localized but not its strength. Only in a
second and independent step when one specifies the sources and
therefore the strength of $G_{lmnp}$ (this is where the boundaries
come in) as given in (\ref{SourceParam}) does the $\kappa^{2/3}$
parameter appear. It has therefore nothing to do with the
structure of the bulk solution and can in particular not imply its
linearization. This becomes very clear when one goes to the limit
in which the $K3\times \mathbf{S}^1/\mathbf{Z}_2$ decompactifies
and for consistency one has to recover from (\ref{6dMetricSol})
the extremal M5 brane solution. Though in this limit $w$ acquires
also a dependence on $y^m$ the crucial point is that only with the
warp-factors of the exact solution is one able to recover
precisely the M5 brane solution \cite{WWarp} whose charge (which
gets matched with the $G_{lmnp}$ flux in $w$) is as well
proportional to $\kappa^{2/3}$. Coming back to our exact solution
(\ref{MetricSol}) describing compactifications down to 4d the
argumentation will be exactly the same with the difference that as
a further consistency check we will not match it to an M5 brane
solution but instead we will show later that upon reducing our
exact 11d background to 5d one recovers precisely the domain wall
solution of \cite{DW}. Again this is only possible by keeping the
full warp-factors as opposed to linearizing them.

Let us now discuss the two different sorts of expansions which
arise in the construction of heterotic M-theory and in obtaining
the linearized background approximation. To this end let us first
consider the $\kappa^{2/3}$ expansion of heterotic M-theory in 11d
prior to any compactification. It is well-known that M-theory in
the absence of any scalar fields does not possess a dimensionless
coupling constant which could serve as a small expansion
parameter. This is the major problem when one aims to establish
M-theory as a theory of supermembranes in analogy to the
perturbative definition of string-theories. In heterotic M-theory
there is however one fundamental scalar modulus, the orbifold
length $L$. So a natural dimensionless expansion parameter is
$\kappa^{2/9}/L$. However, the $\kappa^{2/3}$ expansion of
heterotic M-theory is actually only a formal expansion (see also
the remarks in \cite{HW2}). Namely, whatever the dimensionless
expansion parameter would be, a conventional expansion would
require smaller and smaller higher order terms. This means that
the `higher order' boundary action $S_{bound}$ would have to be
sufficiently suppressed against the `leading order' bulk action
$S_{bulk}$, i.e.~$S_{bulk}\gg S_{bound}$. However, precisely this
is not the case, because what one finds is $S_{bulk}+S_{bound}=0$
(see \cite{CK1} for the case with $\kappa^{2/3}$ level truncated
background and section three below for the case with exact
non-linear background) in accord with a vanishing cosmological
constant at tree-level.

We will see however now that the warp-factor in the exact
non-linear geometry admits an expansion in terms of a small
dimensionless `parameter' for which it is necessary to {\it
distinguish more carefully between the linearized background
approximation and the first order $\kappa^{2/3}$ approximation of
Horava-Witten theory}. To arrive at the linearized geometry namely
requires the further assumption that
\beqa
1\gg \Big|
\sum_i(x^{11}-x_i^{11})\Theta(x^{11}-x_i^{11})\cS_i\Big| \; ,
\label{ParamRestrict}
\eeqa
is satisfied where the `expansion parameter' -- the full
expression inside the absolute values on the rhs -- involves not
only $\kappa^{2/3}$ through $\cS_i$ but $x^{11}$ as well. Once
this additional condition is met we can approximate $e^f$ by $1+f$
and find that $f$ approaches the linearized warp-factor $f_{lin}$
\beqa
f(x^{11})\rightarrow f_{lin}(x^{11}) = - \frac{2}{3}
\sum_i(x^{11}-x_i^{11})\Theta(x^{11}-x_i^{11})\cS_i  \; .
\eeqa
In order to see the precise relation between the exact and the
linearized solution, let us express the exact warp-factor $e^f$
through $f_{lin}$
\beqa
e^{f(x^{11})} = \Big| 1 + \frac{3}{2}f_{lin}(x^{11})
\Big|^{\frac{2}{3}}
\eeqa
which for $\frac{3}{2}|f_{lin}|<1$ can be expanded in terms of the
infinite Binomial series
\beqa
\sum_{n=0}^\infty {2/3 \choose n}\left(\frac{3f_{lin}}{2}\right)^n
\label{Series}
\eeqa
and thus shows explicitly how and which higher order contributions
$f_{lin}^n,\; n\ge 2$ are included in the exact non-linear
solution. As mentioned earlier the {\it inclusion of the
higher-order terms eliminates the stronger bound $\epsilon \ll 1$}
whose duty was to guarantee (via $f_{lin} =
-\frac{2}{3}x^{11}\cS_v =
-\frac{x^{11}}{3\sqrt{2}V_v}\int_{CY}\omega\we S_v$ and the
estimate $\int_{CY} \omega \wedge S_v / V_v \simeq
(\kappa/V_v)^{2/3}$) that higher order corrections $f_{lin}^n,\;
n\ge 2$ to $f_{lin}$ are negligible and the linear background
approximation (\ref{WittenSol}) could be trusted.

Notice that in general the constraint (\ref{ParamRestrict})
imposes not only a constraint on the fluxes but also on the range
of $x^{11}$. Necessarily, in the vicinity of every point where the
warp-factor $e^f$ becomes zero, the approximation
(\ref{ParamRestrict}) and therefore the linearized background are
no longer valid because the `expansion parameter' becomes of
$\cO(1)$ as can be seen from (\ref{WarpFactor}). Consequently,
whenever one has to enter this regime (e.g.~to evaluate the
effective 4d Newton's Constant) one has to use the exact
non-linear background.

\subsection{The Standard Vacua Without M5-Branes}

For clarity and because this gives the simplest vacua, let us
briefly illustrate the geometry for the original Horava-Witten
set-up which includes the visible boundary contribution but omits
additional M5 sources\footnote{\label{hidden flux footnote}The
hidden boundary source is not independent but related to the other
sources through the anomaly cancellation condition.}. Here the
warp-factor becomes (cf.~fig.\ref{LinVac})
\beqa
e^{f(x^{11})} = \left|1 - x^{11}\cS_v
\right|^{\frac{2}{3}}
\label{WarpFacSimple}
\eeqa
while the CY volume becomes a simple parabola
(cf.~fig.\ref{LinVac})
\begin{equation}
  V(x^{11}) = \left( 1-x^{11}\cS_v \right)^2 V_v \; .
\label{VolSimple}
\end{equation}
\setcounter{figure}{0}
\begin{figure}[t]
\begin{center}
\begin{picture}(305,100)(0,-20)
%Volume-Pic
\LongArrow(5,2)(5,100)

\Line(5,80)(35,-10)
\Curve{(5,80)(30,23.75)(55,5)(80,23.75)(105,80)}

\LongArrow(-10,5)(110,5)
\Line(55,3)(55,7)
\Line(3,80)(7,80)
\Line(30,3)(30,7)

\Text(-5,80)[]{$V_v$}
\Text(4,-5)[]{$0$}
\Text(26,-7)[]{$\frac{1}{2}$}
\Text(130,7)[]{$x^{11}\cS_v$}
\Text(55,-7)[]{$1$}
\Text(27,100)[]{$V(x^{11})$}

%WarpFactor-Pic
\LongArrow(205,2)(205,100)
\LongArrow(190,5)(310,5)

\Curve{(205,95)(207.5,91.9744)(210,88.8953)(212.5,85.7585)
(215,82.5596)(217.5,79.2934)(220,75.9536)(222.5,72.5333)
(225,69.0241)(227.5,65.4159)(230,61.6964)(232.5,57.8507)
(235,53.8595)(237.5,49.698)(240,45.3326)(242.5,40.7165)
(245,35.7796)(247.5,30.408)(250,24.3899)(252.5,17.2149) (255,5)}
\Curve{(255,5)(257.5,17.2149)(260,24.3899)(262.5,30.408)
(265,35.7796)(267.5,40.7165)(270,45.3326)(272.5,49.698)
(275,53.8595)(277.5,57.8507)(280,61.6964)(282.5,65.4159)
(285,69.0241)(287.5,72.5333)(290,75.9536)(292.5,79.2934)
(295,82.5596)(297.5,85.7585)(300,88.8953)(302.5,91.9744) (305,95)}

\Curve{(205,95)(255,35)(305,-25)}

\Line(255,3)(255,7)
\Line(203,95)(207,95)

\Text(200,95)[]{$1$}
\Text(204,-5)[]{$0$}
\Text(330,7)[]{$x^{11}\cS_v$}
\Text(255,-7)[]{$1$}
\Text(239,100)[]{$e^f$,\;$1+f_{lin}$}
\end{picture}
\caption{The left figure shows the positive parabolic CY volume
which results from using the exact non-linear background in
comparison to the linearized approximation which is given as the
tangent to the parabola at the location of the visible boundary.
The right figure compares the exact warp-factor $e^f$ (curve with
the peak at the singularity) with the linearized warp-factor
$1+f_{lin}$ (straight line). To trust the linearized
approximations requires $x^{11}\cS_v \ll 1$ which is only valid in
the constrained parameter regime where $\epsilon\ll 1$. This
constraint is not needed when one works with the exact
background.}
\label{LinVac}
\end{center}
\end{figure}
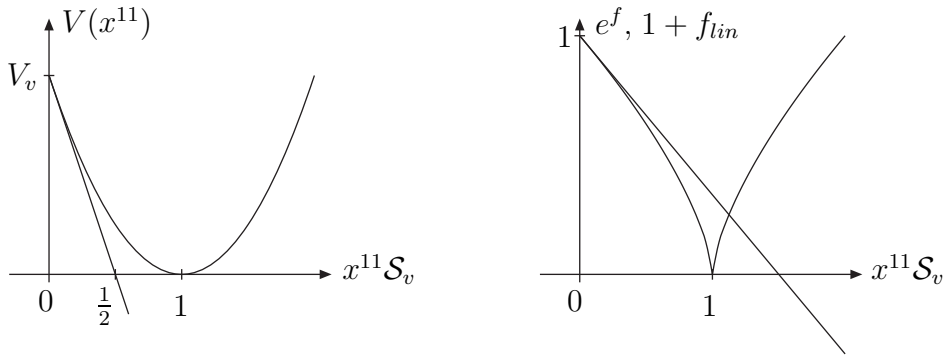
Obviously, at $x^{11}_0=1/\cS_v$ the CY volume and the warp-factor
vanish. Moreover, a chirality change occurs for the hidden
boundary gaugino when the hidden boundary passes this point
(i.e.~when one increases the length modulus $L$ in a movement in
moduli space of heterotic M-theory beyond the value $1/\cS_v$)
combined with a sign-change in the kinetic term of the gaugino.
Under the premise that M-theoretic corrections might resolve the
singularity at $x_0^{11}$ and for completeness we analytically
extend the solution to the regime $x^{11}\ge 1/\cS_v$ depicted
also in fig.\ref{LinVac}. The linearized approximation which
constitutes the first order approximation of a Taylor series
expansion in $x^{11}\cS_v$
\begin{alignat}{3}
f(x^{11}) &= \frac{2}{3}\ln(1 - x^{11}\cS_v)
\simeq - \frac{2}{3}x^{11}\cS_v
= f_{lin}(x^{11})
\label{Lin}    \\
V(x^{11}) &\simeq (1-2x^{11}\cS_v)V_v
\end{alignat}
where
\beqa
\cS_v
= -\frac{1}{8\pi V_v}\left(\frac{\kappa}{4\pi}\right)^{2/3}
\int_{CY}\omega\we(\tr F\we F -\frac{1}{2}\tr R\we R)_v
\eeqa
requires the {\it further} assumption of $x^{11}$ being not too
large to guarantee $x^{11}\cS_v \ll 1$. In the vicinity of the
singular point $x^{11}\lesssim x_0^{11}$ where this requirement is
not fulfilled one has to keep all higher order terms arising from
expanding the logarithm in (\ref{Lin}).

One could think that the kink which is present at $x_0^{11}$ in
the non-linear geometry might already tell us that the Einstein
equations require a source being localized at this point (see
e.g.~\cite{SpaTay} where singularities require the introduction of
effective three-branes in order to balance the Einstein
equations). This however turns out not to be true because a closer
look at the Einstein tensor $E_{MN}=R_{MN}-(1/2)R {\hat g}_{MN}$
for the non-linear background reveals that
\begin{alignat}{3}
E_{\mu\nu} &= -\frac{3}{2}\eta_{\mu\nu}e^{-2f}
\big(f''+(f')^2\big) \\
E_{mn} &= -g_{mn}\big(\frac{1}{2}f''+(f')^2\big) \\
E_{11,11} &= \frac{3}{4}(f')^2
\end{alignat}
which does not lead to a Dirac delta-function
$\delta(1-x^{11}\cS_v)$ which would indicate a localized source
but gives instead a simple power divergence showing again the
presence of the singularity. Roughly speaking, the geometry does
not enforce the localization of a source at $x^{11}_0=1/\cS_v$
because also the first derivative of the non-linear warp-factor
diverges here. If it would stay finite, then the Einstein tensor
would exhibit a delta-function at this point thereby telling us
that some source (basically the hidden boundary because charge
conservation and supersymmetry preservation exclude other choices)
would have to be placed here. One may think of different possible
mechanisms to `resolve' the singularity. Taking into account
perturbative corrections from higher derivative terms like $R^4$
terms \cite{BG}, \cite{Strom2}, \cite{AMTV} leads to a shift of
the CY-volume proportional to the Euler number and an ensuing
effective lower bound on the `quantum' CY-volume. Alternatively,
non-perturbative corrections like M5-instantons wrapping the CY
\cite{GutSpa} become important when the CY-volume goes to zero
classically\footnote{Note that when considering quantum volumes
\cite{Greene} the points where the CY collapses (essentially the
type IIA mirror dual of the type IIB conifold singularity) is not
necessarily the same as the point where a 2-cycle, for example,
collapses (here the development of the CY along the
$x^{11}$-interval is considered (adiabatically) as a movement in
type IIA moduli space); so the discussion of special points in the
interval will be more involved.}. Moreover, we have seen that in
order to avoid hidden gauginos with a wrong sign kinetic term the
hidden boundary should be placed before or at the singularity. If
it is placed at the singularity also gauge instantons of the
hidden $E_8$ gauge group should become relevant \cite{WWarp}.

To summarize, having first order fluxes in Horava-Witten
supergravity does not force one to use the linearized background
approximation as well; rather, on the contrary, one must use the
exact non-linear background with its quadratic volume dependence
as one is approaching the volume's zero position (singularity).
Moreover, the phenomenon of a chirality change at the singularity
appears only in the exact background because only here can one go
beyond this point which was not permitted in the linearized
background.

\section{The Exact Background: 11d Origin of the 5d Domain
Wall Solution}

We will now show that the exact background constitutes the 11d
origin of the 5d domain wall solution discovered in \cite{DW}. The
5d domain wall solution is a supersymmetric solution to the
equations of motion of the effective 5d heterotic M-theory without
the incorporation of further M5 branes. It is well-known that
because it is an exact solution it also incorporates higher order
$\kappa^{2n/3}$ contributions and reduces to the linearized
background dimensionally reduced to 5d upon linear truncation to
$\kappa^{2/3}$ order. It is thus natural to speculate that our 11d
exact background taken without M5 branes might give the exact 5d
domain wall solution upon reduction from 11d to 5d.

To verify that this is indeed the case, let us write our 11d line
element (\ref{MetricSol}) in terms of the CY volume
(\ref{VolSimple})
\beqa
ds^2 = \left(\frac{V(x^{11})}{V_v}\right)^{-2/3} ds_5^2
+ \left(\frac{V(x^{11})}{V_v}\right)^{1/3} g_{mn}(y) dy^m dy^n
\eeqa
where
\beqa
ds_5^2 = \left(\frac{V(x^{11})}{V_v}\right)^{1/3} \eta_{\mu\nu}
dx^\mu dx^\nu + \left(\frac{V(x^{11})}{V_v}\right) dx^{11} dx^{11}
\; .
\eeqa
\begin{figure}[t]
\begin{center}
\begin{picture}(305,120)(0,-20)

\LongArrow(0,77)(0,15)
\Text(40,45)[]{\small dim.~reduction}
\Text(0,90)[]{11d Linearized Background}
\Text(0,0)[]{5d Linearized Background}

\LongArrow(230,90)(90,90)
\Text(160,80)[]{\small linear truncation}

\LongArrow(310,77)(310,15)
\Text(270,45)[]{\small dim.~reduction}
\Text(310,90)[]{11d Exact Background}
\Text(310,0)[]{5d Domain Wall Solution}

\LongArrow(230,0)(90,0)
\Text(160,10)[]{\small linear truncation}

\end{picture}
\caption{The diagram of connections between various heterotic
M-theory backgrounds in 11d and 5d.}
\label{Net}
\end{center}
\end{figure}
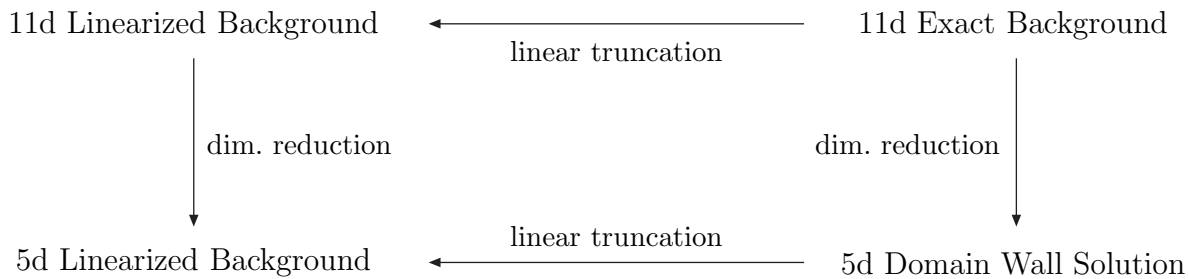
Next let us perform a coordinate transformation from the interval
coordinate $x^{11}$ to another interval coordinate $z$ defined
through\footnote{We are working in the downstairs picture. In the
upstairs picture one would have to extend this identification to
$(1-\frac{2}{3}|z|\cS_v)^3 \equiv V(|z|) = \frac{V(|x^{11}|)}{V_v}
= (1-|x^{11}|\cS_v)^2$ and would then recover the 5d domain wall
solution as well.}
\beqa
(1-\frac{2}{3} z \cS_v)^3 \equiv V(z) = \frac{V(x^{11})}{V_v} =
(1-x^{11}\cS_v)^2 \; .
\label{Trafo}
\eeqa
Such a transformation is always allowed as it leaves the action
and the Bianchi-identity for $G$ invariant. It implies that
\beqa
\left(\frac{V(x^{11})}{V_v}\right)^{1/2} dx^{11}
= V^{2/3}(z) dz
\eeqa
and therefore the 11d line-element expressed in terms of the new
orbifold coordinate $z$ becomes
\beqa
ds^2 = V^{-2/3}(z) ds_5^2
+ V^{1/3}(z) g_{mn}(y) dy^m dy^n
\label{11dMetric}
\eeqa
with
\beqa
ds_5^2 = V^{1/3}(z) \eta_{\mu\nu} dx^\mu dx^\nu
+ V^{4/3}(z) dz^2 \; .
\label{5dMetric}
\eeqa
We next observe that according to the conventions of \cite{DW}
(\ref{11dMetric}) has now the right form such that a dimensional
reduction of this 11d metric down to 5d results in an effective 5d
metric given by $ds_5^2$. Hence the reduction of our exact 11d
background delivers a 5d metric solution given by (\ref{5dMetric})
which together with the explicit $z$ dependence of $V(z)$ given in
(\ref{Trafo}) turns out to be precisely the 5d domain wall
solution presented in \cite{DW} (up to some constants which have
been trivially absorbed). Notice that beyond the agreement of the
geometries also the type of flux being switched on agrees. Both
our 11d background (\ref{MetricSol}) as well as the 5d domain wall
solution have just the $G_{lmnp}$ flux component being switched on
while $G_{lmn11}$ is zero in both cases. We have thus established
the diagram of relations between various background solutions of
heterotic M-theory as given in fig.\ref{Net}.

\section{Vanishing of the Cosmological Constant in the
Non-Linear Background}

Let us now determine the tree-level vacuum energy of heterotic
M-theory in the exact non-linear background. For simplicity we
will do this for the standard vacua without additional M5 branes.
Indeed we will find a vanishing cosmological constant in
accordance with the fact that this background is supersymmetry
preserving. While this had been checked in \cite{CK1} only at the
$\kappa^{2/3}$ truncation level of the exact background, i.e.~for
the linearized background, we will prove it here for the exact
non-linear background. The result is then another manifestation of
the fact that the non-linear background, though it involves higher
order $\kappa^{2n/3},\; n \ge 2$ field-theory corrections, is
protected by supersymmetry and does not mingle with the unknown
higher order $\kappa^{2n/3},\; n \ge 2 $ corrections of heterotic
M-theory. These as we have pointed out would require a deviation
from the local field-theory description.

So let us adopt the vacua described in the previous section
without additional M5 branes. The singularity which occurred at
$x_0^{11}=1/\cS_v$ would, if it can be lifted by M-theoretic
corrections, presumably contribute additional energy sources. We
will therefore constrain the analysis to the regime where
$L<1/\cS_v$, i.e.~space ends at the hidden boundary before the
singularity is reached. Only in this section we will work in the
upstairs picture because only here is an integration over
boundary-localized sources unambiguous. That means instead of
working on the 11d orbifold interval
$\mathbf{M}^{11}/\mathbf{Z}_2$ (downstairs picture, employed in
the rest of this paper) we will work on the smooth 11d manifold
$\mathbf{M}^{11}$ together with the $\mathbf{Z}_2$ symmetry acting
on $x^{11}$. In particular the 11d interval
$\mathbf{S}^1/\mathbf{Z}_2$ becomes a circle $\mathbf{S}^1$ with
periodicity $2L$.

Clearly the 11d bulk curvature scalar $R$ will pick up
delta-function localized contributions from both boundaries. The
$\mathbf{Z}_2$ symmetry demands that the metric components
\beqa
{\hat g}_{AB}(-x^{11}) = {\hat g}_{AB}(x^{11}) \; , \qquad {\hat
g}_{11,11}(-x^{11}) = {\hat g}_{11,11}(x^{11})
\eeqa
are both even. So, extending the metric solution (\ref{MetricSol})
from $\mathbf{M}^{11}/\mathbf{Z}_2$ to $\mathbf{M}^{11}$, we get
\beqa
e^{f(x^{11})} = \big( 1-|x^{11}|\cS_v \big)^{2/3}\; , \;\;\; -L
\le x^{11} \le L
\eeqa
which has a kink at the visible boundary (we omitted the absolute
value in the warp-factor (\ref{WarpFacSimple}) due to the
restriction $|x^{11}|\le L < 1/\cS_v$). Consequently $R$ involving
the second derivative of the warp-factor receives a delta-function
contribution. By using the periodicity of the 11d circle one
infers likewise a singular contribution from the hidden boundary.
The contributions to $R$ can therefore be split into a regular and
a singular part
\beqa
R({\hat g}) = Reg + Sing \; .
\label{Aux1}
\eeqa
The determination of the regular piece can be straightforwardly
done from the metric
\beqa
Reg = \Big[ e^{-f}\big( 2f'' + \frac{5}{2}(f')^2
\big)\Big]_{regular}
= -\frac{2}{9}\frac{\cS_v^2}{(1-|x^{11}|\cS_v)^{8/3}} \; .
\label{Aux2}
\eeqa
Because the singular part is more subtle let us derive it by a
different strategy which will allow us at the same time to obtain
a useful relation needed later on. To this end let us contract the
Einstein equations \cite{LOW2} ($I,J,K$ run from $0$ to $10$ and
$A,B,C$ from $0$ to $9$)
\begin{alignat}{3}
R_{MN}-\frac{1}{2}{\hat g}_{MN}R =
&-\frac{1}{24}(4G_{MIJK}{G_N}^{IJK}-\frac{1}{2}{\hat g}_{MN}
G_{IJKL}G^{IJKL})    \notag \\
&-\delta^A_M\delta^B_N
\Big( \delta(x^{11})T^v_{AB}+\delta(x^{11}-L)T^h_{AB} \Big)
\label{Einstein}
\end{alignat}
where the boundary energy-momentum tensors are given
through\footnote{The coefficient $c$ appearing in the boundary
energy momentum tensor would be $c=1$ if one would use the result
of Horava-Witten for the relation between the 10d gauge coupling
and the 11d gravitational coupling. However in \cite{JOC} the
local anomaly cancellation had been reconsidered and a value of
$c=1/(2^{1/3})$ had been found in the upstairs picture ($c=1/2$ in
the downstairs picture). Notice that $c$ drops out in the end in
(\ref{balance}) and thus cannot be determined by demanding a
vanishing cosmological constant.} ($i=v,h$)
\begin{equation}
\sqrt{{\hat g}_{11,11}}\,T^i_{AB} = \frac{c}{2\pi}
\left(\frac{\kappa}{4\pi}\right)^{2/3}
\Big(\tr F^i_{AC} F^{i\,C}_B - \frac{1}{4}{\hat g}_{AB}
\tr(F^i)^2 -\frac{1}{2}\big( \tr R_{AC}{R_B}^C - \frac{1}{4}{\hat
g}_{AB} \tr R^2 \big)\Big)
\label{EMT}
\end{equation}
with ${\hat g}^{MN}$ and obtain
\begin{equation}
-\frac{1}{24}G_{IJKL}G^{IJKL} = 3R - \frac{2}{3}
\Big( \delta(x^{11})T_A^{v\,A}+\delta(x^{11}-L)T_A^{h\,A} \Big) \; .
\label{RiemScal}
\end{equation}
Now, though $G$ itself experiences a sign-jump at each boundary,
$G_{IJKL}G^{IJKL}$ stays regular when crossing a boundary and in
particular does not develop any singularities. Therefore
delta-function singularities appear only on the rhs of
(\ref{RiemScal}) and consequently have to cancel which implies
that (where the index contractions of the gauge and gravitational
field-strengths are done with the {\em unwarped} original CY
metric)
\begin{alignat}{3}
Sing &= \frac{2}{9}
\Big( \delta(x^{11})T_A^{v\,A}+\delta(x^{11}-L)T_A^{h\,A} \Big)
\label{Aux3} \\
&= - \frac{c}{6\pi}
\left(\frac{\kappa}{4\pi}\right)^{\frac{2}{3}}
\frac{1}{(1-x^{11}\cS_v)^{5/3}}
\Big( \delta(x^{11})(\tr F^2 - \frac{1}{2}\tr R^2)_v
+ \delta(x^{11}-L)(\tr F^2 - \frac{1}{2}\tr R^2)_h \Big) \notag
\end{alignat}
Note a relation (which will be used below) which one finds from
(\ref{RiemScal})
\beqa
-\frac{1}{24}G_{IJKL}G^{IJKL} = 3 Reg \; .
\label{Aux4}
\eeqa

Before coming to the main determination of the vacuum energy, let
us briefly present another useful identity. The expression for the
visible boundary source
\begin{equation}
(S_v)_{lmnp} = -\frac{c}{4\sqrt{2}\pi}
\left(\frac{\kappa}{4\pi}\right)^{2/3}
3!\Big(\tr F_{[lm}F_{np]}-\frac{1}{2}\tr R_{[lm} R_{np]}\Big)_v
\end{equation}
together with the relation (following from the supersymmetry
conditions $F_{ab}=F_{\abar\bbar}=0, \; \omega^{lm}F_{lm}=0$,
where again index contractions are performed with the {\em
unwarped} CY metric)
\begin{equation}
\omega^{lm}\omega^{np} \tr F_{[lm}F_{np]} = -\frac{2}{3}\tr
F_{lm}F^{lm}
\end{equation}
and the analogous one for $R_{lm}$ allow us to rewrite the visible
boundary flux $\cS_v$ as
\begin{equation}
\cS_v = \frac{c}{16\pi V_v}\left(\frac{\kappa}{4\pi}\right)^{2/3}
\int_{CY} d^6y \sqrt{g_{CY}}
\Big(\tr F_{lm}F^{lm} - \frac{1}{2}\tr R_{lm}R^{lm} \Big)_v
\label{Aux5}
\end{equation}
(and similarly for $\cS_h$; the contractions are performed with
the {\em unwarped} CY metric). These identities allow us to
express the boundary action in terms of the respective fluxes.

Let us now calculate the vacuum energy by starting with the bulk
action (in the upstairs picture) which by using (\ref{Aux1}),
(\ref{Aux2}), (\ref{Aux3}) and (\ref{Aux4}) becomes\footnote{For
fluxes with support on a CY threefold or a 7-manifold the CS-term
vanishes trivially.}
\begin{alignat}{3}
S_{bulk} &= -\frac{1}{2\kappa^2}\int_{\mathbf{M}^{11}}
\!\!\!\!\!\!\!  d^{11}x \sqrt{-{\hat g}}\,\Big(
R+\frac{1}{24}G_{IJKL}G^{IJKL}+
\frac{\sqrt{2}}{1728}\epsilon^{I_1\hdots I_{11}}
C_{I_1I_2I_3}G_{I_4I_5I_6I_7} G_{I_8I_9I_{10}I_{11}} \Big)\notag \\
&= \frac{1}{2\kappa^2}\int d^{10}x \int_{-L}^L dx^{11}
\sqrt{-\hat{g}}\,\big( 2Reg - Sing \big)  \notag \\
&= \frac{1}{2\kappa^2}\int d^4x\sqrt{-g_4}\,
\Big( \frac{4}{3}V_v\cS_v \big(1-\frac{1}{(1-L\cS_v)^{2/3}}\big)
+\frac{8}{3}V_v \big(\cS_v+\frac{\cS_h}{(1-L\cS_v)^{2/3}}\big)
\Big)
\end{alignat}
where we have used (\ref{Aux5}) and its hidden boundary equivalent
in the second bracket of the third row to convert the boundary
integrals into the flux parameters $\cS_{v,h}$. The anomaly
cancellation condition (\ref{ACC}) thus gives for the bulk
contribution to the vacuum energy
\beqa
S_{bulk} = \frac{1}{2\kappa^2}\int d^4x\sqrt{-g_4}\, 4V_v\cS_v
\Big( 1-\frac{1}{(1-L\cS_v)^{2/3}} \Big)
\eeqa

Next, let us evaluate the contribution from the boundary action
which is given by\footnote{The $\tr R_{lm}R^{lm}$ terms are the
only non-vanishing terms of the complete Gauss-Bonnet combination
which was argued for in \cite{LOW1} on the basis of supersymmetry.
The two other terms, $R_{AB}R^{AB}$ and $R_{10}^2$, vanish because
both the 10d Ricci tensor and the 10d curvature scalar are zero
because of the Ricci-flatness of the CY. }
\beqa
S_{bound} = -\frac{c}{8\pi\kappa^2}
\left(\frac{\kappa}{4\pi}\right)^{2/3}
\sum_{i=v,h} \int_{\mathbf{M}^{10}_i} d^{10}x \sqrt{-{\hat
g}_{10}}
\Big( \tr F_{lm} F^{lm}
- \frac{1}{2}\tr R_{lm}R^{lm} \Big)_i
\eeqa
The index contractions are done with the exact {\em warped}
metric. The evaluation of the boundary action for the non-linear
background and the identity (\ref{Aux5}) plus its counterpart for
the hidden boundary lead, with the anomaly cancellation condition
(\ref{ACC}), to
\beqa
S_{bound} &=& - \frac{1}{2\kappa^2}\int d^4x \sqrt{-g_4}\, 4V_v
\Big( \cS_v + \frac{\cS_h}{(1-L\cS_v)^{2/3}} \Big) \nonumber\\
&=& - \frac{1}{2\kappa^2}\int d^4x \sqrt{-g_4}\, 4V_v\cS_v\Big( 1
- \frac{1}{(1-L\cS_v)^{2/3}} \Big) \; .
\eeqa
Therefore the bulk and boundary contributions to the vacuum energy
precisely cancel
\beqa
\label{balance}
S_{bulk}+S_{bound} = 0 \; .
\eeqa
Thus the cosmological constant vanishes at tree-level. Although
expected from supersymmetry it is non-trivial to get this
remarkable {\em all-orders} result (in field-theoretical
$\kappa^{2/3}$ corrections included in the non-linear background)
in the strongly coupled M-theory computation. The origin is the
relation (\ref{WarpFacSimple}) between the $G$-flux and the warped
geometry whose structure is protected by supersymmetry and tempts
one to believe that it might be possible to encode the
supersymmetry conditions (\ref{FirstSet}), (\ref{SecondSet}) in
terms of likewise supersymmetry protected perfect squares $R({\hat
g})
+ \frac{1}{24}G^2 \sim$ `$(\partial f - G)^2$' in the effective 4d
potential (cf.~for the weakly coupled heterotic string the similar
second order in $\alpha'$ perfect square structure \cite{CCDL}
resulting from the balance between the $H$-flux and the
deformation of the CY geometry (measured by the deviation $dJ\neq
0$ from being K\"ahler) in the 4d tree-level
potential\footnote{The related issue of the superpotential is
discussed in \cite{Supo},\cite{CCDL}.}). To carry this out in
detail would require to incorporate $f''$ into a perfect square as
well. Indeed equ.~(\ref{Aux4}) suggests such a connection; however
note that one is not allowed to use (\ref{Aux4}) if one wants to
deduce that supersymmetry solutions fulfill the equations of
motion because (\ref{Aux4}) relies already on the equations of
motion.

\section{Newton's Constant}

We have seen so far that the $\epsilon\ll 1$ constraint is an
artefact of the linearized background approximation and that by
including non-linear effects flux compactifications of heterotic
M-theory do not require this constraint. If the distribution of
the sources is such that there are no singularities then one is
free to extend $L$ to arbitrary size. In this case a prediction of
the 4d Newton's Constant requires a stabilization mechanism for
$L$. Such a mechanism had been proposed in \cite{CK2} for vacua
including an additional M5 brane of the sort introduced earlier.
It exploited non-perturbative open membrane instantons stretching
from the M5 to each of the boundaries in combination with
$G$-fluxes. We will evaluate the resulting 4d Newton's Constant
for this scenario in the non-linear background and will see that
it comes quite close to its measured value.

If on the other hand a singularity appears at some $x_0^{11}$, one
would have to cut it out in a purely classical supergravity
framework. This would then in a spirit similar to \cite{WWarp}
lead to an upper bound on $L$. However, because we are dealing
with M-theory which extends classical general relativity one
should expect that the singularity which is time-independent will
be lifted by quantum effects or the extended nature of the
M-theory membrane (see e.g.~the discussion in \cite{BG}, or the
review \cite{LMS}). If this resolution mechanism is local,
i.e.~affects only the singularity but not the rest of the
geometry, then at $x^{11}\ge x_0^{11}$ we enter a region in which
the gaugino has the wrong sign for its kinetic term. Unless we can
make sense of this situation it tells us that the hidden boundary
should sit at some $L\le x_0^{11}$. We will then see in this
section that the position $L= x_0^{11}$ is highly favoured by
giving an astonishingly precise value for the 4d Newton's
Constant. We will nevertheless present formulae for the 4d
Newton's Constant which are general enough to cover also the
situation where the hidden boundary sits at some $L\ge x_0^{11}$.

For the background (\ref{MetricSol}) Newton's Constant can be
derived explicitly by integrating out the internal dimensions in
the 11-dimensional Einstein-Hilbert action\footnote{Here we are
working on the orbifold $\mathbf{M}^{11}\!/\mathbf{Z}_2$, i.e.~in
the downstairs picture. Therefore the usual $1/(2\kappa^2)$ factor
multiplying the Einstein-Hilbert action has to be replaced by
$1/\kappa^2$. See footnote three of \cite{HW2}.}
\begin{alignat}{3}
S &= -\frac{1}{\kappa^2}\int d^4x\int d^6y\int_0^L dx^{11}
\sqrt{-{\hat g}^{(11)}}R({\hat g}^{(11)})    \notag \\
&= -\frac{1}{\kappa^2}\int d^4x \sqrt{-g^{(4)}}R(g^{(4)})V_1
\int_0^L dx^{11}|\cF|^5\;\, + \;\,\hdots
\end{alignat}
where the hatted metric includes the warp-factors in
(\ref{MetricSol}) and we have used the relations $R({\hat
g}^{(11)})=e^{-b}R(g^{(4)}) + \hdots$ and $\sqrt{-{\hat
g}^{(11)}}=|\cF|^3\sqrt{-g^{(11)}}$. Notice the absolute value in
the second row which becomes essential if a singularity is present
and one enters with $L$ the regime beyond the singularity. This
absolute value had been neglected in \cite{CK1} and therefore gave
incorrect results in this regime. Comparison with the standard 4d
Einstein-Hilbert action
\beqa
-\frac{1}{16\pi G_4}\int d^4x\sqrt{-g^{(4)}}R(g^{(4)})
\eeqa
gives for Newton's Constant the general formula
\begin{alignat}{3}
G_4 &= \frac{\kappa^2}{16\pi V_v}
\frac{1}{\int_0^L dx^{11}|e^{\frac{5}{2}f}|}  \notag \\
&= \frac{\kappa^2}{16\pi V_v} \frac{1}{\int_0^L
dx^{11}|1-\sum_i\cS_i\times(x^{11}-x^{11}_i)
\Theta(x^{11}-x^{11}_i)|^{\frac{5}{3}}} \; .
\end{alignat}
Similarly by integrating out the internal dimensions in the gauge
field actions
\beqa
-\frac{1}{8\pi(4\pi\kappa^2)^{\frac{2}{3}}}\int d^{10}x
\sqrt{-{\hat g}^{(10)}} {\hat g}^{AC}{\hat g}^{BD}\tr F_{AB}
F_{CD}\Big|_{v,h}
\eeqa
comparison with the standard 4d gauge-field action
\beqa
-\frac{1}{16\pi\alpha_{v,h}}\int d^{4}x \sqrt{-g^{(4)}}
g^{\mu\rho}g^{\nu\sigma} \tr F_{\mu\nu} F_{\rho\sigma} \Big|_{v,h}
\eeqa
gives the following values for the respective visible and hidden
gauge coupling constants
\beqa
\alpha_v = \frac{(4\pi\kappa^2)^{\frac{2}{3}}}{2V_v} \; , \qquad
\alpha_h = \frac{(4\pi\kappa^2)^{\frac{2}{3}}}{2V(x^{11}=L)} \; .
\eeqa

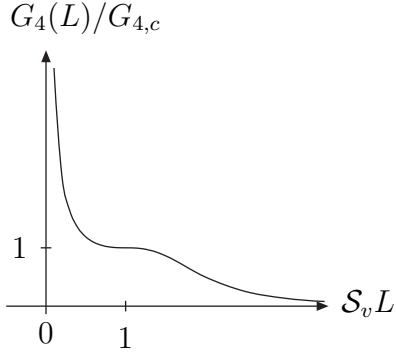
\begin{figure}[t]
\begin{center}
\begin{picture}(100,100)(0,-5)
\LongArrow(5,2)(5,100)

\Curve{(8,94.82)(11,54.06)(14,40.85)(17,34.57)(20,31.11)
(23,29.09)(26,27.93)(29,27.31)(32,27.05)(35,27.00)
(38,26.95)(41,26.70)(47,25.24)(50,24.01)(65,16)(80,10.57)
(95,7.99)(110,6.76)}

\LongArrow(-10,5)(110,5)
\Line(35,3)(35,7)
\Line(3,27)(7,27)

\Text(-5,27)[]{$1$}
\Text(5,-5)[]{$0$}
\Text(127,6)[]{$\cS_v L$}
\Text(35,-7)[]{$1$}
\Text(20,114)[]{$G_4(L)/G_{4,c}$}
\end{picture}
\caption{The 4d Newton's Constant $G_4(L)$ is shown evaluated in
the exact non-linear background (\ref{MetricSol}). $G_{4,c} =
G_4(L
= 1/\cS_v)$ is the value at the point where the CY volume
vanishes. Because in this background the CY volume stays
non-negative for all values of the orbifold size $L$ likewise
$G_4$ stays positive for all $L$.}
\label{Newton}
\end{center}
\end{figure}

Let us evaluate the formula for Newton's Constant first for the
pure Horava-Witten set-up where only boundary sources are present
and second for the situation with an additional M5 brane parallel
to the boundaries which is of relevance for the stabilization
proposal of \cite{CK2}.$^{\ref{hidden flux footnote}}$ For the
first case, we find
\beqa
G_4(L) = \frac{\kappa^2}{6\pi V_v}\frac{\cS_v} {(1 - \sign
(1-\cS_v L) | 1-\cS_v L |^{8/3} )} = \frac{G_{4,c}} {(1 - \sign
(1-\cS_v L) | 1-\cS_v L |^{8/3})}
\eeqa
where
\beqa
G_{4,c} = G_4(L=1/\cS_v) = \frac{\alpha_v^2 V_v}{24\pi^3}
\Big(\frac{4\pi}{\kappa}\Big)^{\frac{2}{3}}\cS_v
= -\frac{\alpha_v^2}{24\pi^2} \int_{CY}\omega \we
\frac{(\tr F\we F - \frac{1}{2}\tr R\we R)_v}{8\pi^2}
\label{GNCrit}
\eeqa
The dependence of $G_4$ on $L$ is depicted in fig.\ref{Newton}. We
see first of all that $G_4$ is a monotonously decreasing function
which stays {\it positive} for all values of $L$. This is in
contrast to the linearized background evaluation where $G_4$
became negative for too large $L$ as a result of the negative
metric problem. However, as discussed before one should rather
place the hidden boundary before the singularity at
$x_0^{11}=1/\cS_v$ to avoid inconsistencies with the hidden
boundary gaugino's kinetic term.
The same would apply if one works exclusively in the classical
supergravity framework and thus has to cut out the singularity.
Then the upper bound on $L\le 1/\cS_v$ results in a lower bound on
$G_4$ which is
\beqa
G_4(L) \ge G_{4,c} \; .
\label{GNCritBound}
\eeqa
Let us now determine this lower bound. Again by estimating the
value of the integral in (\ref{GNCrit}) by $-V_v^{1/3}\simeq
-M_{GUT}^{-2}$ and using $M_{GUT}=3\times 10^{16}$\,GeV together
with $\alpha_v =1/25$ we obtain
\beqa
G_{4,c} = (1.2\times 10^{19}\,\text{GeV})^{-2}
\eeqa
which is in excellent agreement with the observed value
$G_4^{\text{exp}}=(1.22\times 10^{19}\,\text{GeV})^{-2}$.
Actually, the linearized approach \cite{WWarp}
gave\footnote{Already in \cite{WWarp} the lower bound on $G_4$
which was found in the linearized approach came extremely close to
the measured value. By extrapolating the linearized CY volume down
to the point $x^{11}=L_c$ where the linearized CY volume vanishes
and requiring that $L\le L_c$ it was found that $G_4\ge
(3/2)G_{4,c}=(0.98 \times 10^{19}\,\text{GeV})^{-2}$.} $3/2$ times
this value as a lower bound on $G_4$. The task of predicting the
correct value for $G_4$ in the pure (i.e.~without additional
sources) Horava-Witten theory thus becomes the task to find a
mechanism for its stabilization at the orbifold length $L =
1/\cS_v$. Actually the stabilization mechanism of \cite{CK2} gives
exactly\footnote{The orbifold length which was denoted $R\rho$ in
\cite{CK2} gets stabilized at $R=V_1/r_v$ with $V_1v$, $r_v$ the
normalized visible boundary CY volume resp.~the visible boundary
flux ($\rho,v$ were dimensionful normalization parameters). One
has to translate that relation through
$\cS_v=(\frac{r_v}{V_1\rho})_{there}$,
$\cS_{M5}=-(\frac{r_{M5}}{V_1\rho})_{there}$, $L=(R\rho)_{there}$,
$V_v=(V_1v)_{there}$ to the notation employed here giving the
relation $L=1/\cS_v$.} a stabilized $L$ at $L=1/\cS_v$. However,
it employs an additional parallel M5 brane source which changes
the geometry and brings us to our next case. One shouldn't be too
surprised though that the scale is set by the flux as any minimum
of the moduli potential corresponds to a certain length scale and
there are only $\kappa$ (or $\alpha'$ in string-theory) and the
length scale associated to the background value of the fluxes
(essentially given by the compactification scale) available at a
fundamental level. Thus it is essentially the proportionality
constant which has to be determined\footnote{Recently, it was
argued that also in string-theory a stabilization of the dilaton
$g_s$ (which corresponds to $L^{2/3}$ here) at a scale $g_s\simeq
1/Q_{RR}$ set by the RR flux quantum numbers $Q_{RR}$ can be
achieved. See e.g.~the remarks in \cite{FS}.}.

So, let us now consider the situation with an additional parallel
M5 brane which is 4d spacetime filling and wraps an internal
holomorphic 2-cycle $\Sigma$ of the CY (for some background
material concerning the inclusion of M5 branes see
e.g.~\cite{MPS},\cite{M5}). It is located at $x^{11}_{5}$ along
the orbifold interval. With two flux sources present, $\cS_v$ from
the visible boundary and $\cS_5$ from the M5-brane (note that the
hidden boundary source is not independent due to (\ref{ACC})), the
4d Newton's Constant becomes in this case
\beqa
G_4 = \frac{\kappa^2\cS_v}{6\pi V_v F(x^{11}_{5},\cS_v,\cS_{5})}
= \frac{G_{4,c}}{F(x^{11}_{5},\cS_v,\cS_{5})}
\; .
\eeqa
If we define the combined flux parameter $\cS_{v,5} =
(\cS_v+\cS_{5})/(1-\cS_v x^{11}_{5})$, then in the general case in
which $\cS_{v,5}\ne 0$ we find
\begin{alignat}{3}
F(x^{11}_{5},\cS_v,\cS_{5}) &= \frac{\cS_v}{\cS_{v,5}} |1-\cS_v
x^{11}_{5}|^{5/3}
\Bigl( 1- \sign (1-\cS_{v,5}(L-x^{11}_{5}))
|1-\cS_{v,5}(L-x^{11}_{5})|^{8/3}
\Bigr)
\notag \\
&+
\Bigl( 1 - \sign (1-\cS_{v} x^{11}_{5}) |1-\cS_{v}x^{11}_{5}|^{8/3}
\Bigr)
\end{alignat}
For the special case where the fluxes from the visible boundary
and the M5 are equal but opposite (notice from (\ref{HWM5Flux})
that $\cS_{M5}$ is negative for a positively oriented 2-cycle
$\Sigma$ while $\cS_v$ as given in (\ref{HWBoundFlux}) will in
general be positive \cite{WWarp}) i.e.~for the case with
$\cS_{v,5}=0$ one gets
\begin{alignat}{3}
F(x^{11}_{5},\cS_v,-\cS_v) &=  \frac{8}{3}\cS_v (L-x^{11}_{5})
|1-\cS_v
x^{11}_{5}|^{5/3} \notag \\
&+
\Bigl( 1 - \sign (1-\cS_{v} x^{11}_{5}) |1-\cS_{v}x^{11}_{5}|^{8/3}
\Bigr) \; .
\end{alignat}
This latter case appears in the aforementioned stabilization
mechanism of \cite{CK2} where the M5 brane is stabilized at
position $x^{11}_5=L/2$. Actually, the evaluation of the 4d
Newton's Constant with the stabilized value of $L=1/\cS_v$ is not
possible in the linearized background because due to the negative
CY problem it gives a non-physical negative valued Newton's
Constant as well. It is therefore interesting to see that with the
exact non-linear background where these problems are nonexistent
we obtain with $L=1/\cS_v$ and $x^{11}_5=L/2$ in the scenario of
\cite{CK2} an $F(x^{11}_5,\cS_v,-\cS_v)=1.26$ which with the same
values for $M_{GUT}^{-2}$ and $\alpha_v$ as before leads to the
prediction
\beqa
G_4 = (1.35\times 10^{19}\,\text{GeV})^{-2} \; .
\eeqa
Given the uncertainty in estimating the integral appearing in
(\ref{GNCritBound}) by $-V_v^{1/3}$ the stabilization proposal's
result looks very promising on the way to predict 4d parameters
from M-theory.

The full generalization of \cite{CK2},\cite{CK4} to the non-linear
background which requires the reduction of heterotic M-theory over
the non-linear background is under way and we hope to report on
this soon. However, it is clear that in order to predict the right
value for Newton's Constant one would need a stabilization
mechanism which is capable of giving an
$F(x_5^{11},\cS_v,\cS_5)\simeq 1$.

\bigskip
The work of A.K.~has been supported by the National Science
Foundation under Grant Number PHY-0099544. A.K.~wants to thank
M.~Becker, K.~Becker and M.~Luty for discussions related to this
work.

 \newcommand{\zpc}[3]{{\sl Z. Phys.} {\bf C\,#1} (#2) #3}
 \newcommand{\npb}[3]{{\sl Nucl. Phys.} {\bf B\,#1} (#2) #3}
 \newcommand{\plb}[3]{{\sl Phys. Lett.} {\bf B\,#1} (#2) #3}
 \newcommand{\prd}[3]{{\sl Phys. Rev.} {\bf D\,#1} (#2) #3}
 \newcommand{\prb}[3]{{\sl Phys. Rev.} {\bf B\,#1} (#2) #3}
 \newcommand{\pr}[3]{{\sl Phys. Rev.} {\bf #1} (#2) #3}
 \newcommand{\prl}[3]{{\sl Phys. Rev. Lett.} {\bf #1} (#2) #3}
 \newcommand{\jhep}[3]{{\sl JHEP} {\bf #1} (#2) #3}
 \newcommand{\cqg}[3]{{\sl Class. Quant. Grav.} {\bf #1} (#2) #3}
 \newcommand{\prep}[3]{{\sl Phys. Rep.} {\bf #1} (#2) #3}
 \newcommand{\fp}[3]{{\sl Fortschr. Phys.} {\bf #1} (#2) #3}
 \newcommand{\nc}[3]{{\sl Nuovo Cimento} {\bf #1} (#2) #3}
 \newcommand{\nca}[3]{{\sl Nuovo Cimento} {\bf A\,#1} (#2) #3}
 \newcommand{\lnc}[3]{{\sl Lett. Nuovo Cimento} {\bf #1} (#2) #3}
 \newcommand{\ijmpa}[3]{{\sl Int. J. Mod. Phys.} {\bf A\,#1} (#2) #3}
 \newcommand{\rmp}[3]{{\sl Rev. Mod. Phys.} {\bf #1} (#2) #3}
 \newcommand{\ptp}[3]{{\sl Prog. Theor. Phys.} {\bf #1} (#2) #3}
 \newcommand{\sjnp}[3]{{\sl Sov. J. Nucl. Phys.} {\bf #1} (#2) #3}
 \newcommand{\sjpn}[3]{{\sl Sov. J. Particles \& Nuclei} {\bf #1} (#2) #3}
 \newcommand{\splir}[3]{{\sl Sov. Phys. Leb. Inst. Rep.} {\bf #1} (#2) #3}
 \newcommand{\tmf}[3]{{\sl Teor. Mat. Fiz.} {\bf #1} (#2) #3}
 \newcommand{\jcp}[3]{{\sl J. Comp. Phys.} {\bf #1} (#2) #3}
 \newcommand{\cpc}[3]{{\sl Comp. Phys. Commun.} {\bf #1} (#2) #3}
 \newcommand{\mpla}[3]{{\sl Mod. Phys. Lett.} {\bf A\,#1} (#2) #3}
 \newcommand{\cmp}[3]{{\sl Comm. Math. Phys.} {\bf #1} (#2) #3}
 \newcommand{\jmp}[3]{{\sl J. Math. Phys.} {\bf #1} (#2) #3}
 \newcommand{\pa}[3]{{\sl Physica} {\bf A\,#1} (#2) #3}
 \newcommand{\nim}[3]{{\sl Nucl. Instr. Meth.} {\bf #1} (#2) #3}
 \newcommand{\el}[3]{{\sl Europhysics Letters} {\bf #1} (#2) #3}
 \newcommand{\aop}[3]{{\sl Ann. of Phys.} {\bf #1} (#2) #3}
 \newcommand{\jetp}[3]{{\sl JETP} {\bf #1} (#2) #3}
 \newcommand{\jetpl}[3]{{\sl JETP Lett.} {\bf #1} (#2) #3}
 \newcommand{\acpp}[3]{{\sl Acta Physica Polonica} {\bf #1} (#2) #3}
 \newcommand{\sci}[3]{{\sl Science} {\bf #1} (#2) #3}
 \newcommand{\vj}[4]{{\sl #1~}{\bf #2} (#3) #4}
 \newcommand{\ej}[3]{{\bf #1} (#2) #3}
 \newcommand{\vjs}[2]{{\sl #1~}{\bf #2}}
 \newcommand{\hepph}[1]{{\sl hep--ph/}{#1}}
 \newcommand{\desy}[1]{{\sl DESY-Report~}{#1}}

\bibliographystyle{plain}

\end{document}